\documentclass[aps,prd,groupedaddress,twocolumn,nofootinbib,superscriptaddress]{revtex4-1}
\newcommand{\be}{\begin{eqnarray}}
\newcommand{\ee}{\end{eqnarray}}
\usepackage{fancyhdr}
\usepackage{amsmath}
\usepackage{mathrsfs}
\usepackage{float}
\usepackage{xcolor}
\usepackage{graphicx, epsfig, amssymb} 
\usepackage{amsmath, amsfonts}
\usepackage{bm} 
\usepackage{tensor}
\usepackage{soul}
\usepackage{hyperref}

\begin{document}

\title{Effects of the shape of curvature peaks on the size of primordial black holes}

\author{Albert Escriv\`a}
\email{albert.escriva@fqa.ub.edu}
\affiliation{Institut de Ci\`encies del Cosmos, Universitat de Barcelona, Mart\'i i Franqu\`es 1, 08028 Barcelona, Spain}
\affiliation{Departament de F\'isica Qu\`antica i Astrof\'isica, Facultat de F\'isica, Universitat de Barcelona, Mart\'i i Franqu\`es 1, 08028 Barcelona, Spain}
\author{Antonio Enea Romano}
\email{antonio.enea.romano@cern.ch}
\affiliation{Theoretical Physics Department, CERN, CH-1211 Geneva 23, Switzerland}
\affiliation{ICRANet, Piazza della Repubblica 10, I--65122 Pescara, Italy}

\begin{abstract}

We simulate numerically the formation of spherically symmetric primordial black holes (PBHs) seeded by different families of primordial curvature perturbations profiles in a radiation dominated Friedman-Robertson-Walker (FRW) Universe. We have studied the dependency on the curvature profile of the initial mass $M_{\rm BH,i}$  of the PBHs at the time of apparent horizon formation $t_{AH}$, and the final mass $M_{\rm BH,f}$ after the accretion process, using an excision technique, comparing $M_{\rm BH,i}$ to previous analytical estimations obtained using a compensated PBHs model approach. The analytical estimations are in agreement with numerical results, except for large values of the initial perturbation amplitude, when the compensated model is less accurate. The masses $M_{\rm BH,f}$ and $M_{\rm BH,i}$  do not depend only on the shape around the compaction function peak, but on the full profile of the initial curvature perturbation. We also estimate the accretion effects, and for  PBHs with masses relevant for the dark matter abundance, with a final mass equal to the horizon crossing mass, we find $M_{\rm BH,f}\approx 3 M_{\rm BH,i} $.

\end{abstract}

\maketitle

\section{Introduction}\label{sec:intro}

Primordial Black Holes (PBHs) could have been formed in an early period of evolution of our Universe as a consequence of the gravitational collapse of cosmological perturbations \cite{hawking1,hawking2}. Within this hypothesis, it is assumed that PBHs can be generated due to high non-linear rare peaks in the primordial distribution of density perturbations produced during inflation. These perturbations could eventually have collapsed and produced black holes during the radiation domination epoch, or some transitional matter phase \cite{Carr,darkmatter2}.

Currently, there is not a hardbound on the amplitude of the curvature fluctuations at smaller scales than those of the Cosmic Microwave Background Radiation (CMB), leaving open the scenario of having a substantial fraction of the Dark Matter (DM) in the form of PBHs \cite{darkmatter1,Green:2020jor,darkmatter2,darkmatter3,darkmatter4,Atal:2018neu,darkmatter5,darkmatter6,carr_review_new,darkmatter7,darkmatter8,sam_young,Carr:2009jm22,Carr:2020gox22,Ashoorioon:2019xqc,Ashoorioon:2020hln,Fumagalli:2020adf}.

Numerical simulations of the formation of PBHs originated from the collapse of density perturbations has been an active field of research for quite some time \cite{Niemeyer1,Niemeyer2,sasaki,Nakama_2014,refrencia-extra-jaume,mit_pbh,Moradi_2015,musco2005,musco2007,hawke2002}. These numerical simulations are needed to study the gravitational collapse and to determine the initial conditions from which PBHs can be formed, and their masses. A new and more efficient numerical approach has been proposed recently, based on pseudo spectral methods, reproducing previous results in the literature \cite{escriva_solo}.

It was shown in previous studies \cite{Niemeyer1,Niemeyer2,musco2007} that the mass of the PBHs follows a self similar scaling law, but systematic numerical investigations of the PBH size at the apparent horizon formation time and of the effects of  accretion for different profiles of curvature perturbations was not done yet. The importance of the shape of the curvature perturbations was already noticed in \cite{germaniprl,nonlinear,newnicolla,Atal:2019erb,universal1,universal2,Musco:2020jjb,musco2018,Nakama:2013ica,sam}, showing that the threshold for PBH formation \cite{carr75,harada} is a profile dependent quantity.

In \cite{universal1,universal2}, for the first time, it was shown that the threshold for PBH formation mainly depends on the shape around the peak of the compaction function \cite{sasaki} and the equation of state, which was used to build an analytical formula enough accurate for cosmological applications. But a similar dependency for $M_{\rm BH,i}$ and $M_{\rm BH,f}$ has not been demonstrated yet. 
Some theoretical studies have addressed the analytical estimation of an upper bound for $M_{\rm BH,i}$ using a compensated PBH model \cite{size1,size2}, but a systematic numerical investigation exploring the effects of different shapes of the curvature perturbations  was missing. 

The effect of the accretion is known to be negligible for  small $M_{\rm BH,i}$, but it  has not been investigated for the case of the collapse of a perfect fluid for large PBHs, including those with higher probability formation, corresponding to $M_{\rm BH,f} \sim M_{H}$, where $M_{H}$ is the horizon mass at the time of horizon crossing.
A study of the accretion for PBHs formed from the collapse of a massless scalar field was performed in \cite{scalar_harada}, showing that $M_{\rm BH,f}/M_{\rm BH,i}\leq 2$. A similar result was obtained in \cite{deng,vacum_bubles} for the PBH formed from the collapse of domain walls or vacuum bubbles.

Motivated by these open questions, in this work we have addressed these points by performing numerical simulations of the formation process of spherically symmetric PBHs. In section \ref{sec:sharp} we introduce approach used in numerical simulations, in particular, the differential equations we need to solve, boundary conditions, initial conditions, the definition of the threshold for PBH formation, and the condition for the location of the apparent horizon. In section \ref{sec:spectral} we introduce the numerical approach adopted for the simulations, the pseudo-spectral method. In section \ref{sec:aparent_horizon} we make the study the the mass $M_{\rm BH,i}$ and radius $R_{\rm BH,i}$, at the time of apparent horizon formation $t_{AH}$, and we check previous analytical estimations. In section \ref{sec:acretion} we study the accretion process using an excision method, compute the final mass of the PBH, and estimate the accretion effect. 

\raggedbottom

\section{Set up for PBH formation}\label{sec:sharp}

In this section, we resume the suited equations for the evolution of spherically symmetric perturbations leading to PBH formation.

The differential equations that describe the non-linear evolution of a relativistic perfect fluid under spherical symmetry are the Misner-Sharp equations \cite{misnersharp}. In the case of spherical symmetry, the metric of the spacetime can be written as,

\begin{equation}
ds^2 = -A(r,t)^2 dt^2+B(r,t)^2 dr^2 + R(r,t)^2 d\Omega^2,
\label{metricsharp}
\end{equation}

where $d\Omega^{2} = d\theta^2+\sin^2(\theta) d\phi^2$ is the line element of a 2-sphere and $R(r,t)$ is the areal radius. We use units $G_{N}=1$.

The Misner-Sharp mass, includes contributions from the gravitational potential and kinetic energy energies and is introduced as
\begin{equation}
M(r,t) \equiv \int_{0}^{R} 4\pi R^{2} \rho \, R' dr\, ,
\end{equation}
then we can define the $\Gamma$ as:
\begin{equation}
\Gamma = \sqrt{1+U^2-\frac{2 M}{R}}.
\end{equation}

where $U(r,t)$ is the radial component of the four-velocity, which measures the radial velocity of the fluid.

Considering a perfect fluid with an equation of state like $p = w \rho$ (in the case of radiation fluid $w=1/3$), the Misner-Sharp equations written in a convenient way for the numerical implementation are,

\begin{align}
\label{eq:msequations1}
\dot{U} &= -A\left[\frac{w}{1+w}\frac{\Gamma^2}{\rho}\frac{\rho'}{R'} + \frac{M}{R^{2}}+4\pi R w \rho \right], \\
\dot{R} &= A U, \\
\label{eq:msequations2}
\dot{\rho} &= -A \rho (1+w) \left(2\frac{U}{R}+\frac{U'}{R'}\right), \\
\dot{M} &= -4\pi A w \rho U R^{2},
\end{align}

where $(')$ and $(\dot{})$ represents the radial and time derivative respectively. The boundary conditions that should be applied  are $R(r=0,t)=0$, $U(r=0,t)=0$ and $M(r=0,t)=0$. Taking into account spherical symmetry, we have $p'(r=0,t)=0$.

In our work, the Misner-Sharp equations are applied in the cosmological context within a FRW background. In practice, for $r \rightarrow \infty$ we should recover the FRW background, but in a numerical scheme, the grid of the simulation is finite. To avoid possible reflections from pressure waves and to match at the boundary of the grid with the FRW solution, we have implemented $p' (r=r_f,t) = 0$ (where $r_{f}$ is the outer point of the grid). On the other hand, we can solve analytically the lapse function $A(r,t)$ imposing the boundary condition $A(r_f,t) = 1$ to match with the FRW background,

\begin{equation}
A(r,t) = \left(\frac{\rho_{b}(t)}{\rho(r,t)}\right)^{\frac{\omega}{\omega+1}},
\end{equation}
where $\rho_{b}(t) = \rho_{0}(t_{0}/t)^{2}$ is the energy density of the FRW background and $\rho_{0}=3 H_{0}^{2}/8\pi$.

The metric Eq.(\ref{metricsharp}) can be approximated at superhorizon scales and at leading order in gradient expansion by \cite{sasaki}:

\begin{equation}
\label{frwmetric}
ds^2 = -dt^2 + a^2(t) \left[\frac{dr^2}{1-K(r) r^2}+r^2 d\Omega^2 \right].
\end{equation}

The cosmological perturbation will be imprinted in the initial curvature $K(r)$. As was shown in \cite{sasaki}, the mass excess inside a given volume, called compaction function $\mathcal{C}(r)$, is proportional to the product $K(r)r^{2}$ at leading order in gradient expansion \cite{Tanaka:2006zp}. In particular,

\begin{equation}
\mathcal{C}(r,t) = \frac{2 \left[M(r,t)-M_{b}(r,t)\right]}{R(r,t)}.
\label{compactionfunction}
\end{equation}

The peak value of the compaction function, $\mathcal{C}_{\rm max}=\mathcal{C}(r_{\rm m})$, is used as a criteria for PBH formation \cite{refrencia-extra-jaume,sasaki}, where $r_{m}$ is the location of the peak of $\mathcal{C}(r)$. We define the threshold for primordial black hole formation as $\delta_{c} = \mathcal{C}_{c}(r_{\rm m})$ such that a PBH is formed whenever $\bar{\delta}(r_{m})\geq \delta_c$, where $\mathcal{C}_{c}$ is the critical compaction function.

The gradient expansion approximation (or long wavelength approximation) allows to solve Misner-Sharp equations at leading order in $\epsilon \ll 1$ where $\epsilon(t) = R_{H}(t)/a(t)r_{m}$. $r_m$ is the length scale of the perturbation and $R_{H}(t)=1/ H(t)$ is the cosmological horizon. This approach allows us to get the initial conditions for PBH formation in terms of the curvature $K(r)$. They were derived in \cite{musco2007}. 

Alternatively, the metric Eq.\eqref{frwmetric} can be written also in the form putting the curvature fluctuations $\zeta$ outside the 3-metric,

\begin{equation}
\label{eq:zeta_metric}
ds^2 = -dt^2 + a^2(t) e^{2 \zeta(\hat{r})} \left[ d\hat{r}^2+\hat{r}^2 d\Omega^2 \right].
\end{equation}

where the transformation between $K(r)$ and $\zeta(\hat{r})$ was derived in \cite{refrencia-extra-jaume}.

It is useful to know the background quantities: $H(t)=H_{0} t_{0}/t$ , $a(t)= a_{0}(t/t_{0})^{\alpha}$ and $R_{H}(t) = R_{H}(t_{0})(t/t_{0})$ where $a_{0} = a(t_{0})$ , $H_{0}=H(t_{0}) = \alpha /t_{0}$ and $R_{H}(t_{0}) = 1/H_{0}$. We define $\alpha = 2/3(1+\omega)$. We consider a time scale given by $\epsilon(t_{m}) =1$, which leads $t_m = t_{0}(a_{0} r_m/R_H(t_0))^{1/(1-\alpha)}$. The horizon mass $M_H$ at horizon crossing time $t_m$, is given by $M_{H} = \frac{1}{2} r_{m} \left( a_{0} \beta^{\alpha} \right)^{1/(1-\alpha)}$.

We define the amplitude of a cosmological perturbation by the mass excess within a spherical region:
\begin{equation}
\label{massexcess}
\delta(r,t) = \frac{1}{V}\int_{0}^{R} 4\pi R^{2} \frac{\delta \rho}{\rho_{b}} R' dr,
\end{equation}
where $V=4\pi R^{3}/3$ and at leading order in $\epsilon$ gives:
\begin{equation}
\delta(r,t)=\left(\frac{1}{a H r_{m}}\right)^{2} \bar{\delta}(r),
\end{equation}

where $\bar{\delta}(r)=f(w)K(r)r^{2}_{m}$ and  $f(\omega)= 3(1+\omega)/(5+3\omega)$. In the gradient expansion approach, $\mathcal{C}(r,t)\simeq \mathcal{C}(r)=f(\omega) K(r) r^{2} = r^2 \bar{\delta}(r) /r^2_{m}$ \cite{musco2018}, which gives $\mathcal{C}(r_{m}) = \bar{\delta}(r_{m})=\bar{\delta}_{m}$. Due to the above definitions, the value of $r_{m}$ should fulfil,
\begin{equation}
\label{cmax}
K(r_{m})+\frac{r_{m}}{2}K'(r_{m}) = 0.
\end{equation}

The compaction functions remains constant at super horizon scales, but it starts to evolve non-linearly and becomes time dependent once the simulation starts. In our simulations, the formation of a black hole can be inferred by the formation of a trapped surface \cite{penrose}. A trapped surface exist when the expansion $\Theta^{\pm}$ of the two null geodesic congruences $k^{\pm}_{\mu}$ orthogonal to a spherical surface, are negative. Since $\Theta^{\pm} \equiv h^{\mu\nu} \nabla_{\mu}k_{\nu}^{\pm}$ and $k_{\mu}^{\pm} = (A,\pm B,0,0)$,

\begin{equation}
\Theta^{\pm}=\frac{2}{R}(U \pm \Gamma).
\label{eq:congurence}
\end{equation}

In spherical symmetry, the condition for the apparent horizon (AH) is given by $\nabla_{\mu}R\nabla^{\mu} R=2M/R=1$. The AH is a marginally trapped surface, given by the boundary when $\Theta^-<0$ and $\Theta^+=0$. Therefore, to compute the location of AH we need to compute the expansions $\Theta^{\pm}$ numerically at each time steep, and find in what value of $r$ the previous condition is fulfilled.

\section{Numerical technique}\label{sec:spectral}

To perform the numerical simulations, we solve the Misner-Sharp equations using the numerical method developed in \cite{escriva_solo}, which is based on the use of a Pseudo-spectral Chebyshev collocation technique. This method allows us to compute the spatial derivatives with an exponential convergence \cite{spectrallloyd} with the use of the Chebyshev differentiation matrix $D$, which components can be found in \cite{escriva_solo}. The nodes of the Chebyshev grid where we compute the derivatives at given at $x_{k}=\cos(k\pi /N_{\rm cheb})$, where $k=0,1,..,N_{\rm cheb}$ and $N_{\rm cheb}$ is the number of points on the grid. We use an explicit Runge-Kutta method of four-order for the time evaluation.

Already in \cite{universal2}, a substantial improvement regarding efficiency and accuracy was made by the use of composite Chebyshev grids: split the domain in several subdomains to make in each subdomain the desired
Chebyshev grid. Explicitly, our domain is divided into $M$ subdomains given by $\Omega_{l} =[r_{l},r_{l+1}]$ with $l=0,1...,M$. Since the Chebyshev nodes are defined in $[-1,1]$, we also perform a mapping between the spectral and the physical domain for each Chebyshev grid. In particular, we use a linear mapping for each subdomain defined as:
\begin{equation}
 \tilde{x}_{k,l} = \frac{r_{l+1}+r_{l}}{2}+\frac{r_{l+1}-r_{l}}{2}x_{k,l} ,
\end{equation}
where $\tilde{x}_{k,l}$ are the new Chebyshev points re-scaled to the subdomain $\Omega_{l}$. In the same way, the Chebyshev differentiation matrix is re-scaled using the chain rule:
\begin{equation}
 \tilde{D_{l}} = \frac{2}{r_{l+1}-r_{l}} D_{l}.
\end{equation}

On the subdomains, we compute the spatial derivatives using the Chebyshev differentiation matrix $\tilde{D_{l}}$ associated to each subdomain.

Some boundary conditions should be supplied across the different $\Omega_l$s to perform correctly the time evolution. The approach that we have used is the one of \cite{Teukolsky}. We compute the time derivatives of each field at the boundaries between the subdomains. Then, the incoming fields derivative is replaced by the time derivatives of the outgoing fields from the neighbouring domain. Using an analysis of the characteristics of the field, we checked that only the density field is directed outwards, and the others are incoming.

Once an AH is formed, a singularity arises, preventing it from following the numerical simulation. A numerical technique to avoid this and allows us to follow the accretion process from the FRW background is called excision \cite{Teukolsky}. The main idea of excision is that nothing inside the event horizon can affect the physics outside. In our case the excision technique follows the motion of the apparent horizon computing numerically Eq.\eqref{eq:congurence} at each time steep, using a cubic spline interpolator. The method of excision that we use here was used already in \cite{escriva_solo} with only one Chebyshev grid. In this work, we have generalized it with several grid subdomains, which has allowed us to increase the precision and therefore allows more stable long term simulations.

Finally, to test that we are correctly solving Einstein equations at any time, we compute the $L_{2}$ norm of the Hamiltonian constraint equation $M' = 4 \pi  \rho R^{2} R'$ at each time step.

\begin{equation} 
 \mid\mid \mathcal{H} \mid\mid_{2} \equiv \frac{1}{N_{\rm cheb}}\sqrt{\sum_{k} \Big|  \frac{M_{k}'/R_k'}{4 \pi \rho_{k} R_{k}^{2}}-1 \Big|^2},
\label{eq:constraint}
\end{equation}

\section{Apparent horizon formation}\label{sec:aparent_horizon}

In the first part of our work we have computed the size of the PBH at the time $t = t_{AH}$ of  formation of the apparent horizon (AH), when  $2M(r_{AH},t_{AH})=R(r_{AH},t_{AH})$ and $r_{AH}$ is radial coordinate of the AH. 
We run simulations for  different families of initial curvature profiles  \cite{universal1, universal2},

\begin{align}
 \label{basis_pol}
 K_{\rm b}(r) &= \frac{\mathcal{C}(r_m)}{f(w)r_m^2}\frac{1 + 1/q}{1+\frac{1}{q}\left(\frac{r}{r_{m}}\right)^{2(q+1)}}, \\
 \label{eq:lamda}
 K_{\rm exp}(r) &= \frac{\mathcal{C}(r_m)}{f(w) r_m^2}\,
\left(\frac{r}{r_{m}}\right)^{2\lambda}\,
e^{\frac{(1+\lambda)^{2}}{q}\left(1 - \left(\frac{r}{r_{m}}\right)^{\frac{2q}{1+\lambda}}\right)} . \
\end{align}

 Eq.\eqref{basis_pol} has been shown to be a complete basis in the sense that allows to obtain all the possible threshold values $\delta_{c}$ in terms of $q$, which for radiation corresponds to the interval $0.4 \leq \delta_{c} \leq 2/3$ \cite{universal1}. The parameter $q$  is a dimensionless measure of the curvature of $\mathcal{C}(r)$ at its maximum defined as
 \begin{equation}  
q = \frac{-r_{m}^{2} \mathcal{C}''(r_{m})}{4 \mathcal{C}(r_{m})}.
\end{equation}
As  shown in \cite{universal1, universal2}, the threshold for PBH formation only depends on $q$ and the equation of state.

\begin{figure}[!htbp]
\centering
\includegraphics[width=0.7\linewidth]{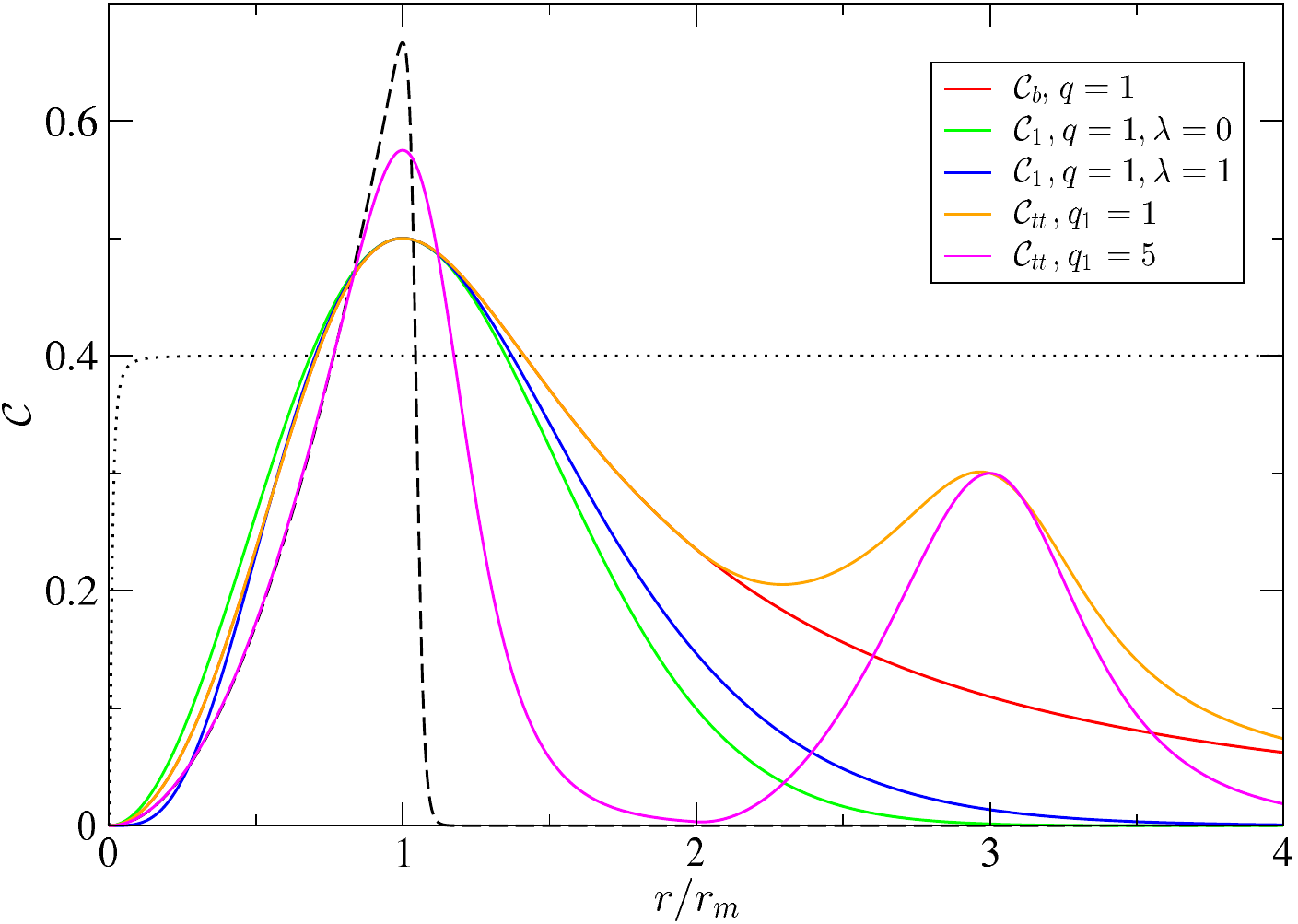} 
\includegraphics[width=0.7\linewidth]{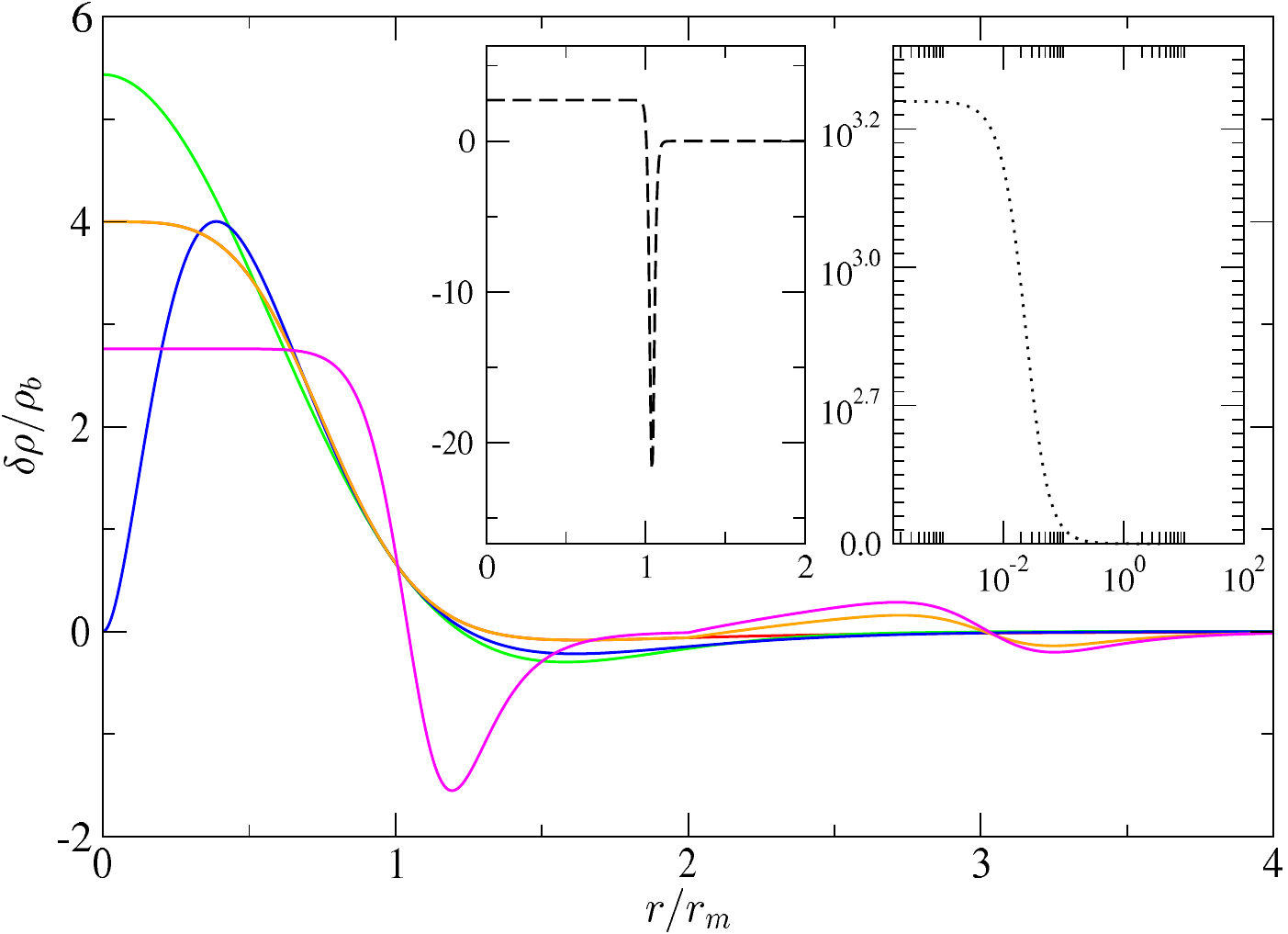} 
\includegraphics[width=0.7\linewidth]{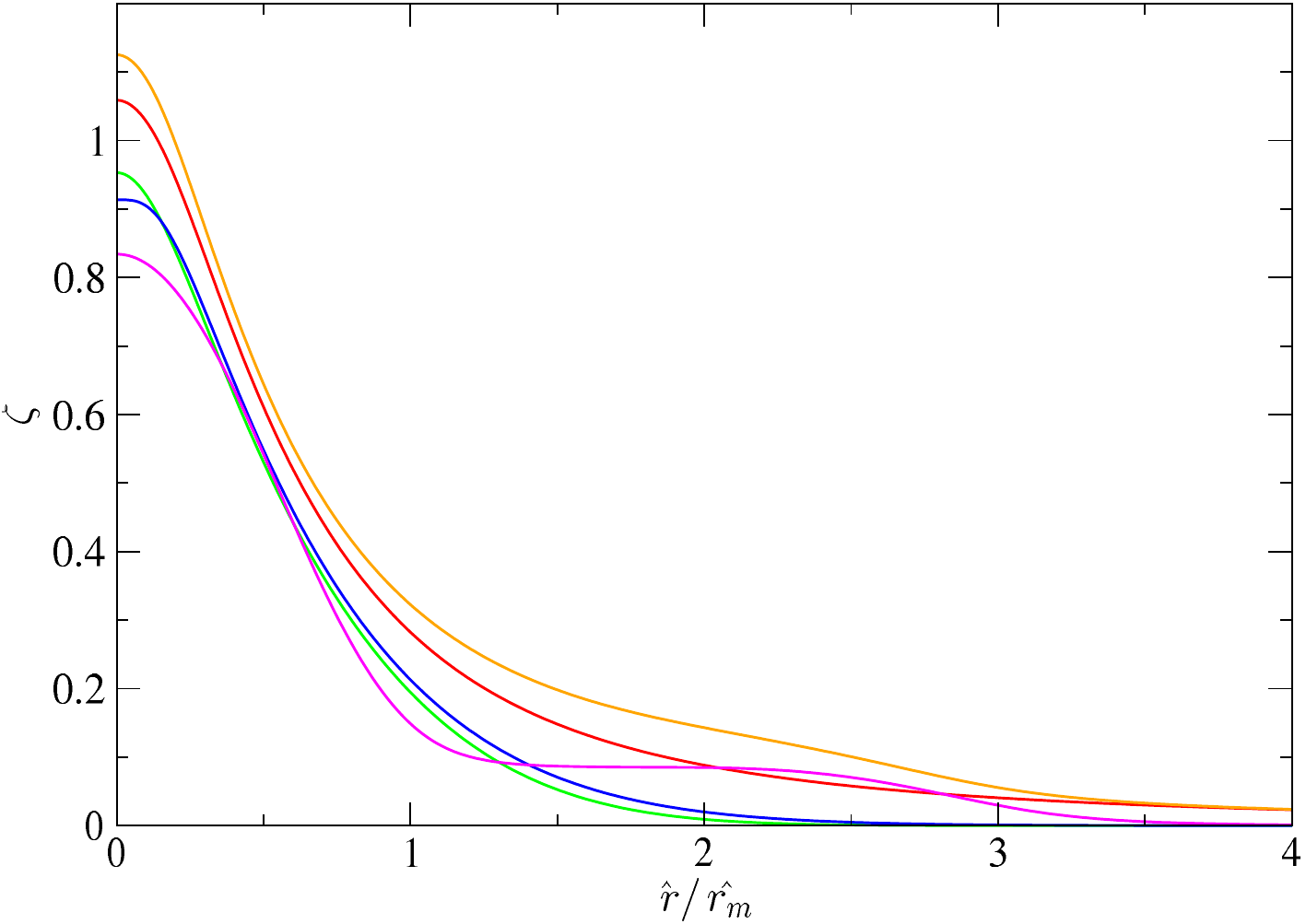} 
\caption{The profiles in Eq.\eqref{basis_pol}, Eq.\eqref{eq:lamda} and Eq.\eqref{eq:dos_torres} for $\mathcal{C}(r)$, $\delta \rho(r)/ \rho_{b}$ and $\zeta(\tilde{r})$ are plotted as function of $r$ for $\delta=\delta_{c}$. The dashed black line corresponds $q \rightarrow \infty$, and the dotted black line to $q=0$. The parameters used for $\mathcal{C}_{tt}(r)$ are $\delta_{1}=\delta_{c}(q_{1})$, $q_{2}=3$, $r_{m1}=r_{m2}=1$, $r_{j}=2r_{m1}$, $\mathcal{C}_{tt(\rm peak,2)}=0.3$, with the corresponding  $\delta_{2}$ obtained from  Eq.\eqref{eq:pico2} using the previous values, and $q_{1}=1$ (orange) and $q_{1}=5$ (violet).}
\label{fig:profiles}
\end{figure}

Different profiles of the two families are plotted in Fig.(\ref{fig:profiles}). For both families when $q \gg 1$ the peak of the compaction function is sharp, while when  $q \ll 1$ the peaks is broad.

\begin{figure}[!htbp]
\centering
\includegraphics[width=1\linewidth]{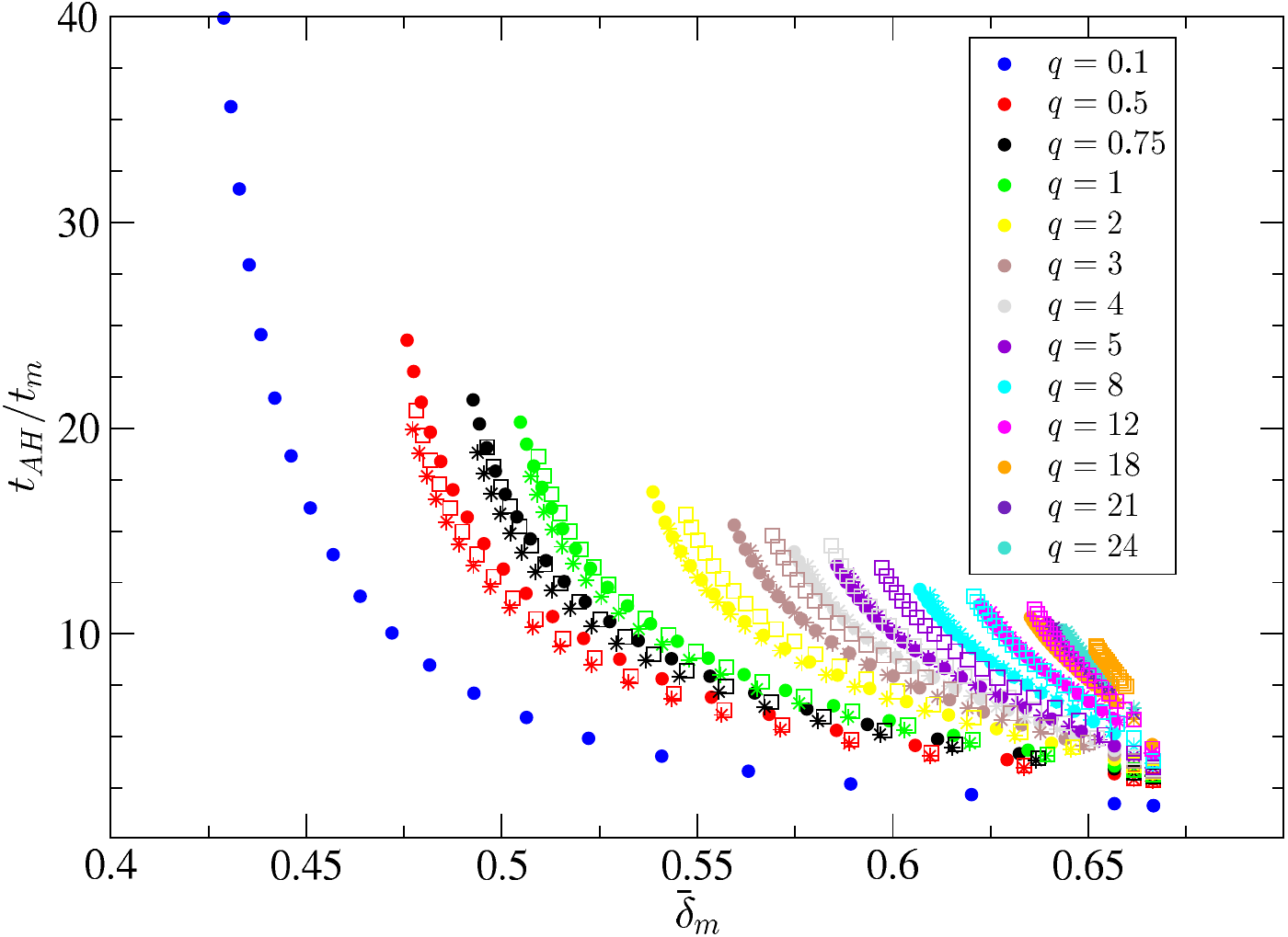} 
\caption{The ratio $t_{AH}/t_{H}$ is plotted as a function of $\bar{\delta}_{m}$ for different values of $q$. As expected, the minimum value of $\bar{\delta}_{m}$, i.e. $\delta_c$, decreases as $q$ decreases. Circle corresponds to Eq.\eqref{basis_pol}, star to Eq.\eqref{eq:lamda} with $\lambda=0$ and square to Eq.\eqref{eq:lamda} with $\lambda=1$.}
\label{fig:time_colapse}
\end{figure}

In Fig.(\ref{fig:time_colapse}) we plot the apparent horizon formation time $t_{AH}$ for different profiles, showing that $t_{AH}$ decreases when $\bar{\delta}_{m}$ is higher since the initial amplitude of the perturbation is much larger than the critical value $\delta_{c}$, and therefore it collapses faster. On the contrary $t_{AH}$ is  large when $\bar{\delta}_{m}$ is close to the critical value. 
This behaviour is the same for different families of profiles.

\begin{figure}[!htbp]
\centering
\includegraphics[width=1.0\linewidth]{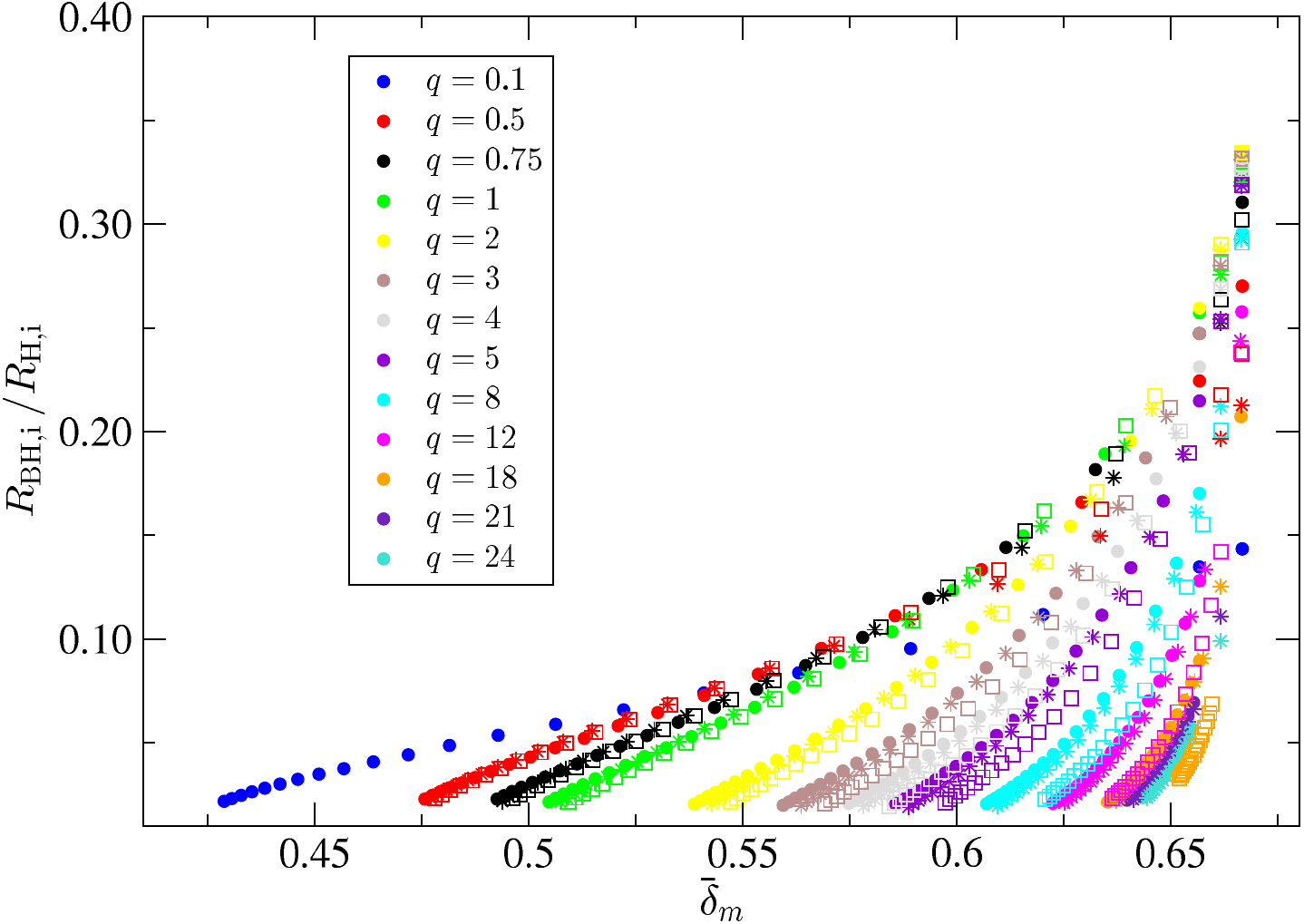} 
\caption{The ratio $R_{\rm BH,i}/R_{\rm H,i}$ is plotted for different values of $q$. As expected, the minimum value of $\bar{\delta}_{m}$, i.e. $\delta_c$, decreases as $q$ decreases. Circle corresponds to Eq.\eqref{basis_pol}, star to Eq.\eqref{eq:lamda} with $\lambda=0$ and square to Eq.\eqref{eq:lamda} with $\lambda=1$.}
\label{fig:radius}
\end{figure}

\begin{figure}[!htbp]
\centering
\includegraphics[width=1.0\linewidth]{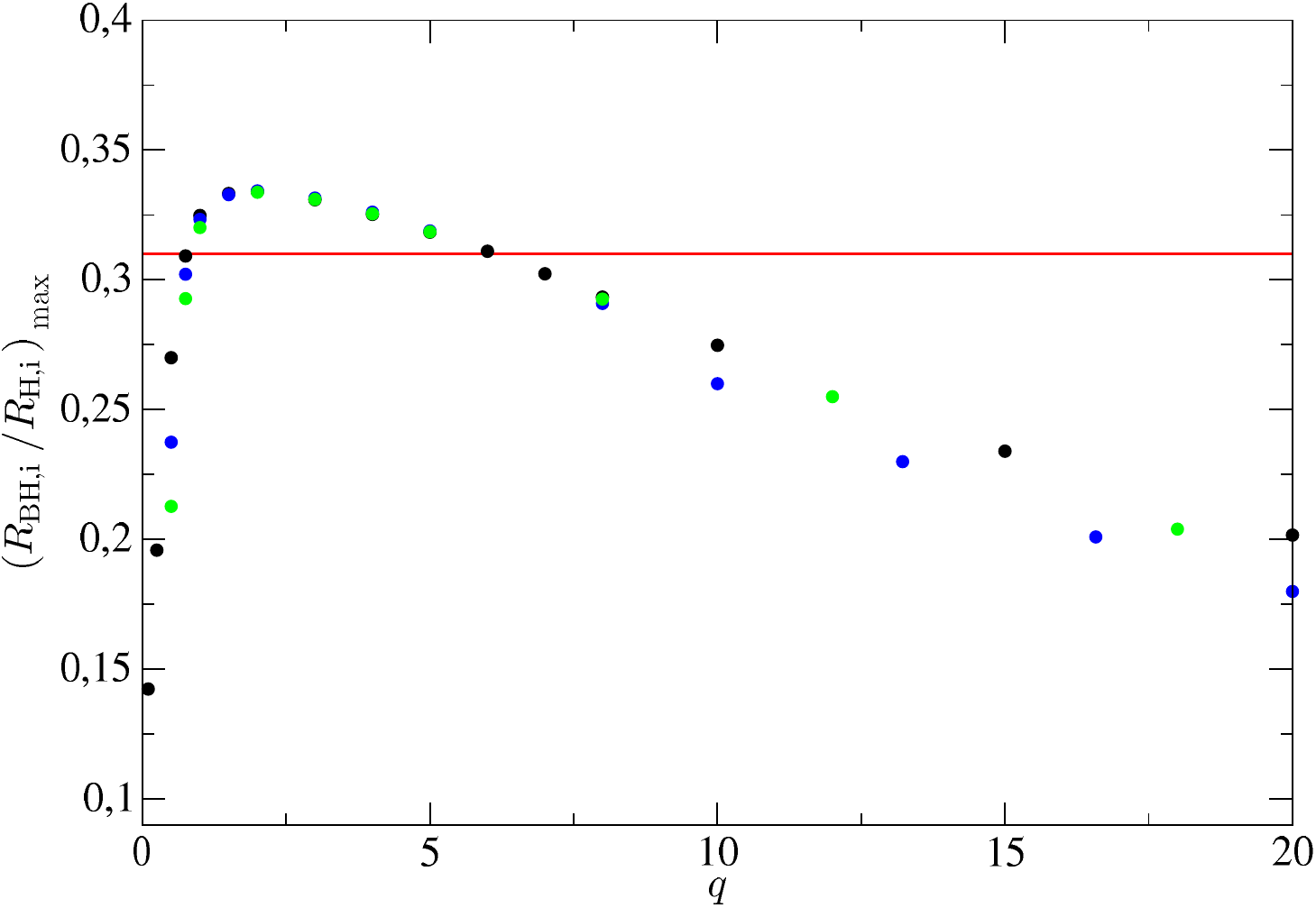} 
\caption{The ratio $R_{\rm BH,i}/R_{\rm H,i}$ is plotted as function of $q$ for $\bar{\delta}_{m} = \delta_{\rm max}-10^{-5}$, using the  profiles in Eq.\eqref{basis_pol} (black), in Eq.\eqref{eq:lamda} with $\lambda=0$ (green) and  in Eq.\eqref{eq:lamda} with $\lambda=1$ (blue). The red line corresponds to the analytical estimation of the upper bound obtained in \cite{size1}.}
\label{fig:bound_R}
\end{figure}

The ratio between the areal radius of the PBH $R_{\rm BH,i}=R(r_{AH},t_{AH})$ and the Hubble radius at  $t_{AH}$, $R_{H,i}$, is plotted in Fig.(\ref{fig:radius}). As expected, for all  PBHs  $R_{\rm BH,i}<R_{H,i}$, because the perturbations collapse after re-entering the the cosmological horizon. In \cite{size1} it was derived an analytical formula for the upper bound of $R_{\rm BH,i}/R_{H,i}$
\begin{equation}
\left(\frac{R_{\rm BH,i}}{R_{H,i}}\right)_{\rm max} = \left(\frac{2}{1+3w}\right)^{3}\left[\frac{3(1+w)}{2(1+\sqrt{w}}\right]^{\frac{3(1+w)}{1+3w}}w^{3/2} \,,
\label{bound}
\end{equation}
, which is approximately confirmed by our numerical results, giving $\approx 0.31$ in the case of radiation,  except for very large values of $\bar{\delta}_{m}$ as shown in Fig.(\ref{fig:bound_R}).

\begin{figure}[!htbp]
\centering
\includegraphics[width=0.8\linewidth]{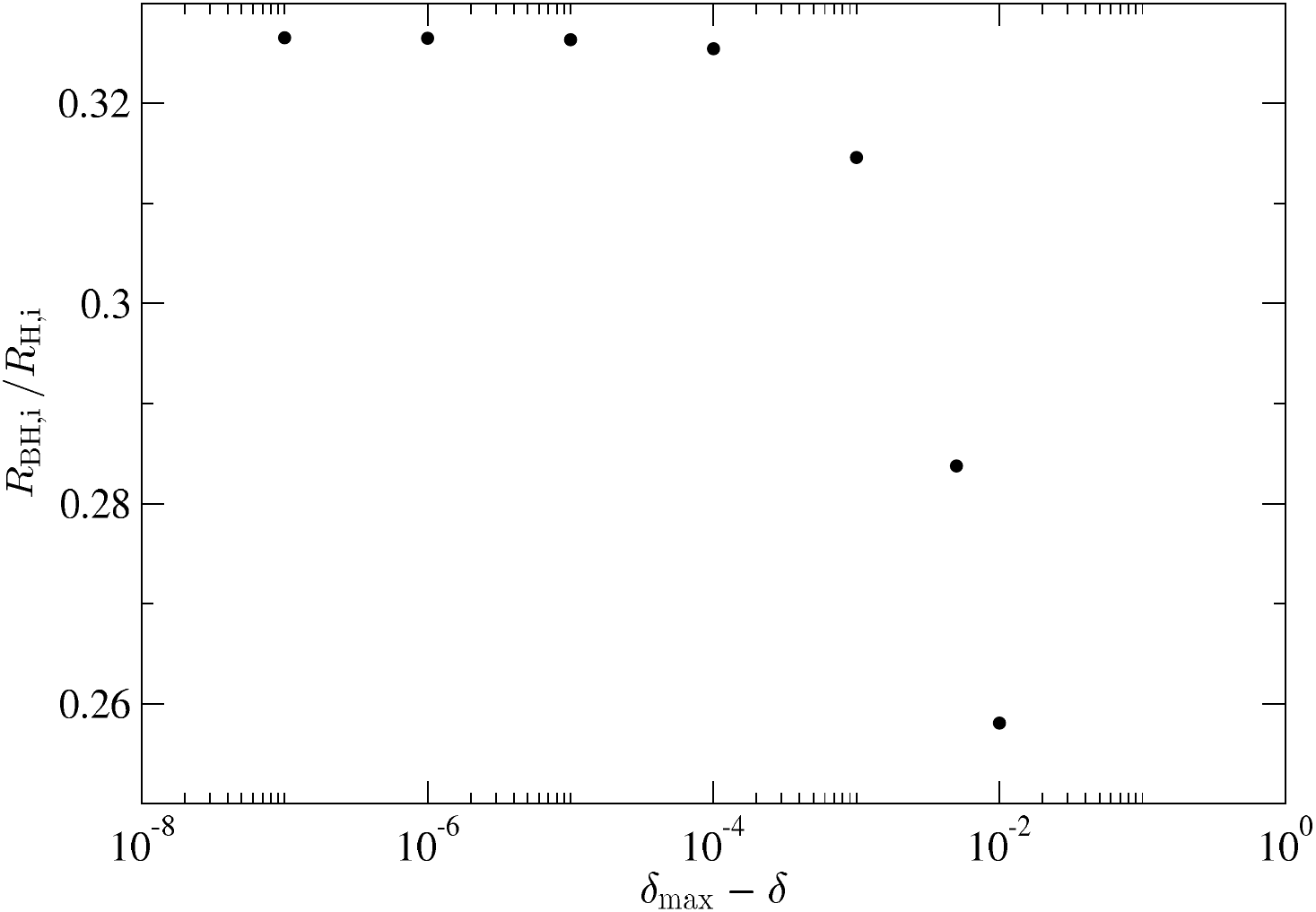} 
\includegraphics[width=0.8\linewidth]{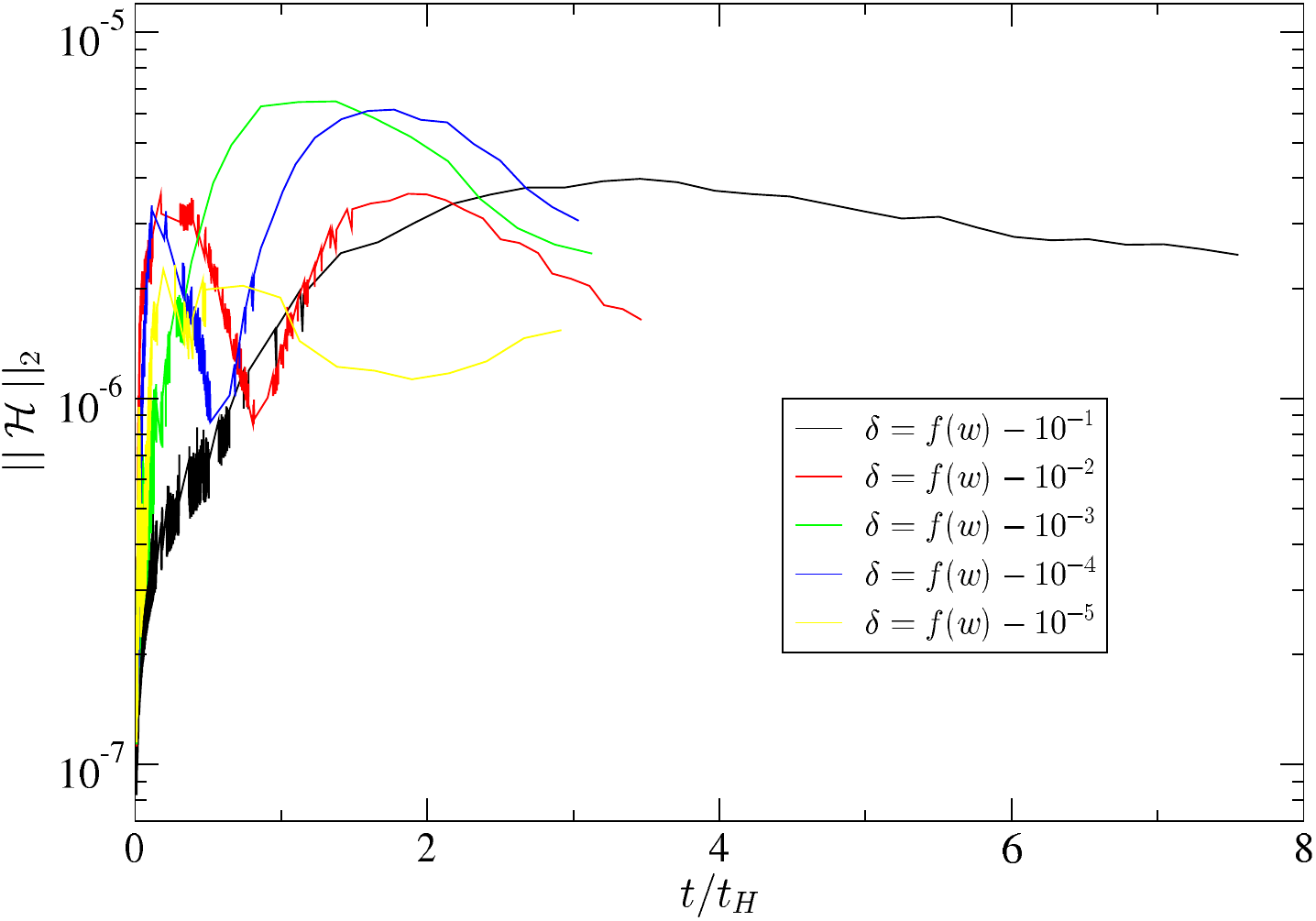} 
\caption{Top: The ratio $R_{\rm BH,i}/R_{\rm H , i}$ is plotted as a function of $\bar{\delta}_{m}$ near the  maximum value $\delta_{max} = f(w)$. Bottom: The time evolution of the Hamiltonian constraint is plotted for the profiles in Eq.\eqref{basis_pol} with $q=1$, and for different values of $\bar{\delta}_{m}$.}
\label{fig:sat}
\end{figure}

Such an analytical bound \cite{size1} was obtained considering a compensated PBH model where the black hole horizon is contained within a perturbed region, surrounded by a FRW background. 
In Fig.(\ref{fig:radius}) we plot the ratio  $R_{\rm BH,i}/R_{H,i}$ for different initial curvature profiles and $\bar{\delta}_{m}$. The upper bound in Eq.\eqref{bound} is satisfied for most of cases except for certain values of $q$ and when $ \bar{\delta}_{m}$ approached the maximum value $\delta_{\rm max}=2/3$ , since when $\bar{\delta}_{m}$ is much greater than the critical value, the ratio can exceed  substantially the bound. This is shown in more details in Fig.(\ref{fig:bound_R}) and Fig.(\ref{fig:sat}), where we have compared the analytically computed bound of the ratio with its numerical calculation.

In this cases the formation time $t_{\rm AH}$ is smaller because the PBH is formed soon after the perturbation crosses the cosmological horizon, and for this reason the ratio $R_{\rm BH,i}/R_{H,i}$ is larger.

\begin{figure}[!htbp]
\centering
\includegraphics[width=1.0\linewidth]{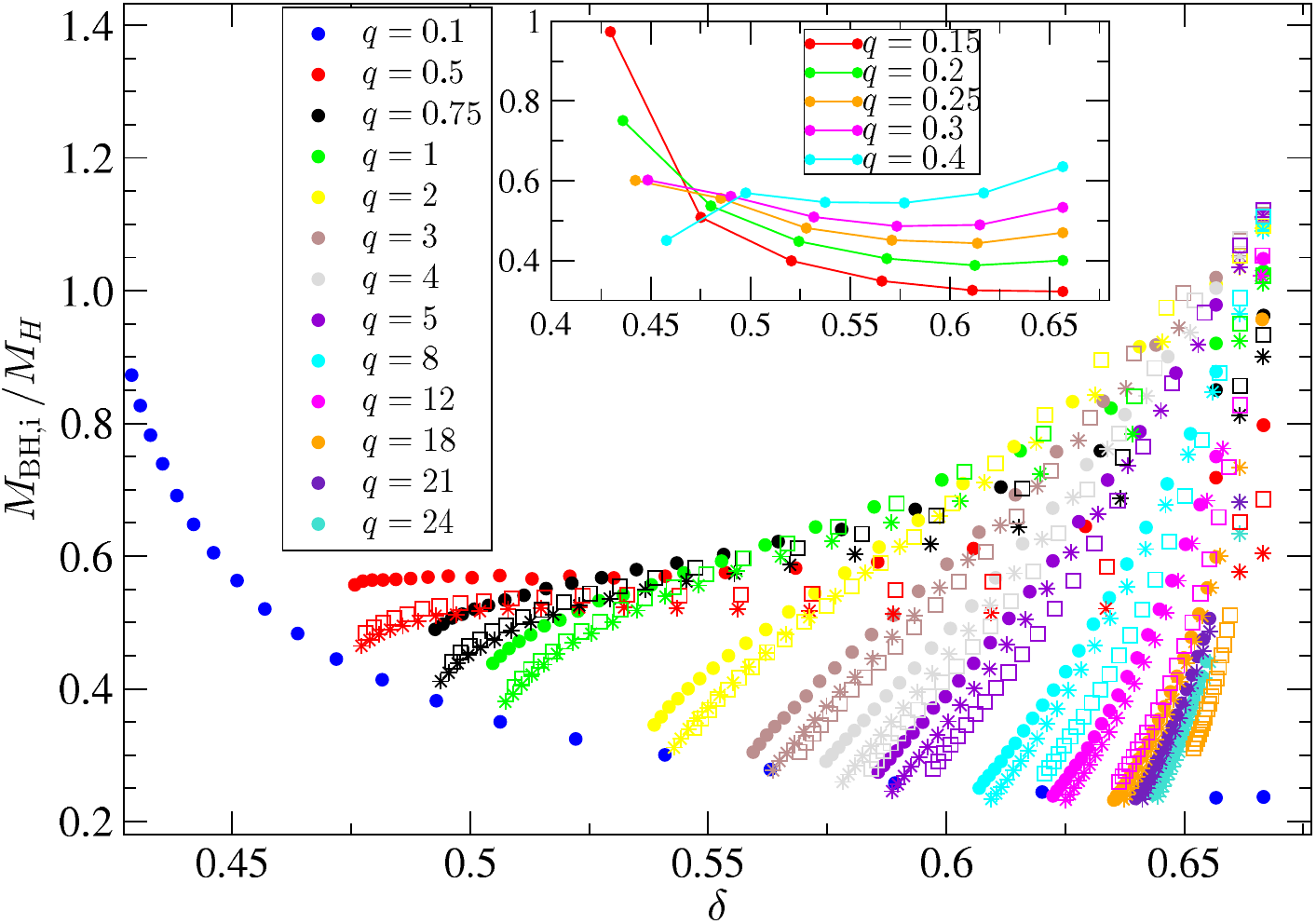} 
\caption{The ratio $M_{\rm BH,i}/M_{\rm H}$ is plotted for different values of $q$. As expected, the minimum value of $\bar{\delta}_{m}$, i.e. $\delta_c$, decreases as $q$ decreases. Circles corresponds to Eq.\eqref{basis_pol}, stars to Eq.\eqref{eq:lamda} with $\lambda=0$ and squares to Eq.\eqref{eq:lamda} with $\lambda=1$.}
\label{fig:mass_pbh}
\end{figure}

{\raggedright

The ratio  $M_{\rm BH,i}/M_H$ is plotted in Fig.(\ref{fig:mass_pbh}) for different initial conditions. As it can be seen, the mass $M_{\rm BH,i}$ is sensitive to  $\bar{\delta}_{m}$ and perturbations with sufficiently large $\bar{\delta}_{m}$ can form black holes with $M_{\rm BH,i}>M_H$.} For small $q$, $M_{\rm BH,i}$ decrease as  $\bar{\delta}_{m}$  increases, since in this case the perturbation collapse faster due to the smaller pressure gradients in comparison when larger $q$, and therefore $M_{\rm BH,i}/M_H$ is smaller. 

\begin{figure}[!htbp]
\centering
\includegraphics[width=1.0\linewidth]{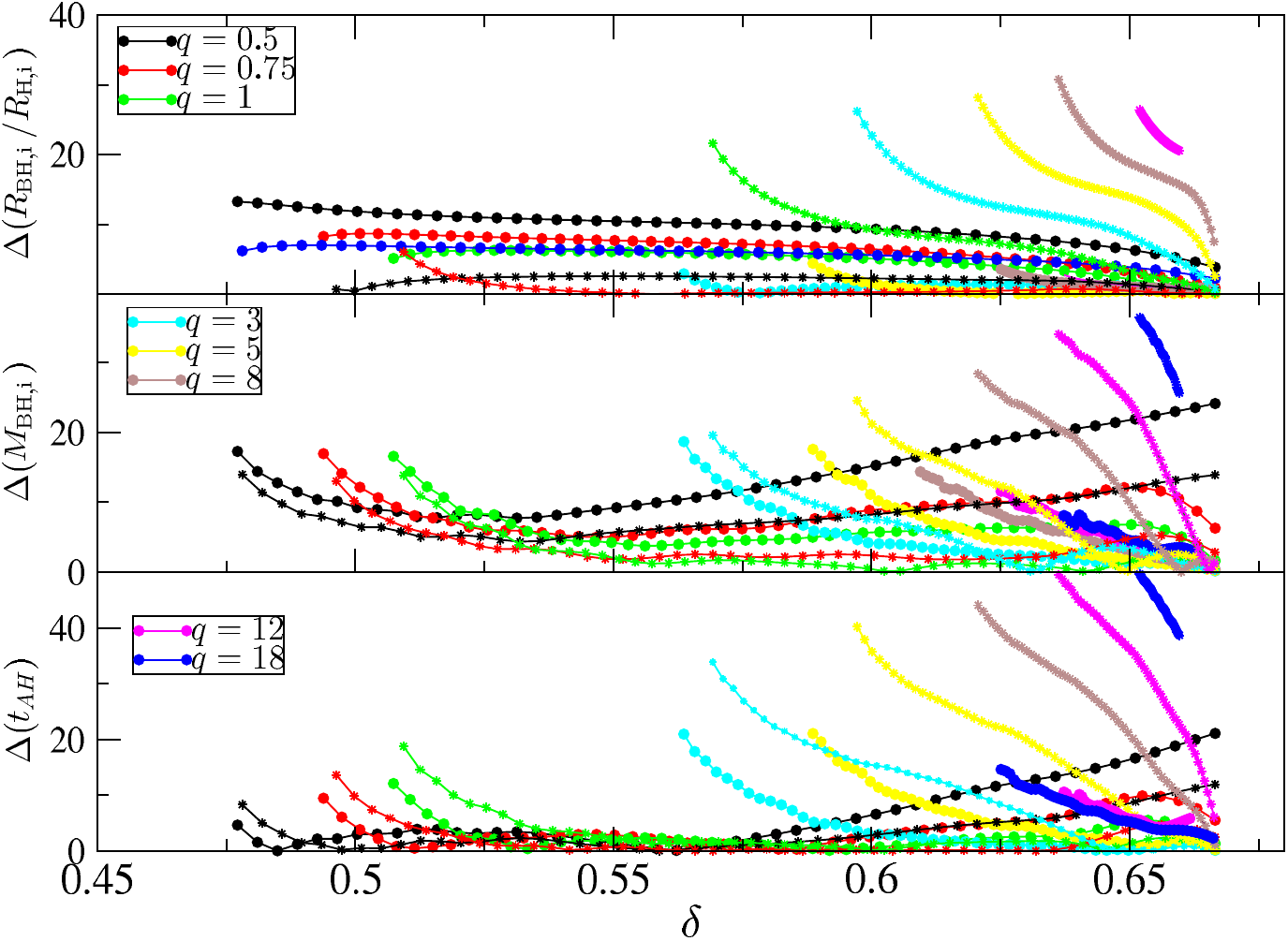} 
\caption{The absolute value of relative percentual difference  between different profile families is plotted for the quantities $R_{\rm BH,i}/R_{H,i}$ (top), $M_{\rm BH,i}$ (middle) and $t_{AH}$ (bottom). The circle points corresponds to the comparison between Eq.\eqref{basis_pol} and Eq.\eqref{eq:lamda} with $\lambda=0$, and square points with q.\eqref{basis_pol} and Eq.\eqref{eq:lamda} with $\lambda=1$.}
\label{fig:deviation}
\end{figure}

A comparison between $t_{AH}$, $R_{\rm BH,i}/R_{H,i}$, $M_{\rm BH,i}$  for different profiles with the same $q$ is shown in Fig.(\ref{fig:deviation}). It is shown there that these quantities are not only q dependent, they depend on the specific details of the profiles considered. The dynamical time scale where the $\delta_{c}$ is determined happens in a time $t_{m} < t \ll t_{AH}$, but the size of the PBH at $t_{AH}$ is determined at later much times, where a substantial part of the profile is involved during the collapse until the formation of the AH.

\section{Effect of the accretion}\label{sec:acretion}

After the formation of the AH it follows a process of accretion that increases the size of the BH until a stationary state with final mass $M_{\rm BH,f}$. 

It has been shown  \cite{acreation1,acreation2,acreation3} that at sufficiently late times of the BH evolution, the mass satisfy this equation
\begin{equation}
\label{Mt}
\frac{dM_{\rm BH}}{dt} = 4 \pi F R^2_{\rm BH} \rho_{b}(t)\ \,,
\end{equation}
where $F$ is the accretion rate constant and it is usually numerically found to be of order $O(1)$, for example in \cite{deng} $F \approx 3.5$. In our case we find $F \in [3.2,3.8]$ in terms of the different profiles, so in agreement with previous results.

The analytical solution of Eq.\eqref{Mt}  during radiation domination is
\begin{equation}
\label{eq:acretationformula}
M_{\rm BH}(t) = \frac{1}{\frac{1}{M_{a}}+\frac{3}{2}F\left(\frac{1}{t}-\frac{1}{t_{a} }\right)}\ ,
\end{equation}
where $M_a$ and $t_a$ define the initial conditions imposed to solve it.

As in \cite{escriva_solo}, we will find $F$ by fitting the numerical evolution of the mass with the formula in Eq.\eqref{eq:acretationformula}. We check that the fit is accurately performed, giving an standard deviation $s_{d}$ of $s_{d}(M_{a}) = 10^{-4.2}$, $s_{d}(t_{a}) = 10^{-4.1}$ and $s_{d}(F) = 10^{-3.6}$. The variance gives $\sigma_{max} \approx 10^{-2}$. Moreover, we expect that the approximation of Eq.\eqref{eq:acretationformula} is valid when $ \Psi = \dot{M}/H M <1$, i.e the increment of the PBH mass respect the Hubble scale. We use a range of numerical values where is fulfilled that $\Psi \lesssim 0.1$ as in \cite{escriva_solo} to make the fit.

Once the best fit parameters  have been determined the final PBH mass is obtained as the asymptotic future limit,

\begin{equation}
\label{eq:massfinalfinal}
M_{\rm BH,f} =\lim_{t \to \infty} M_{\rm BH}(t) = \left(\frac{1}{M_{a}}-\frac{3 F}{2 t_{a}}\right)^{-1} \ .
\end{equation}

\begin{figure}[!htbp]
\centering
\includegraphics[width=1\linewidth]{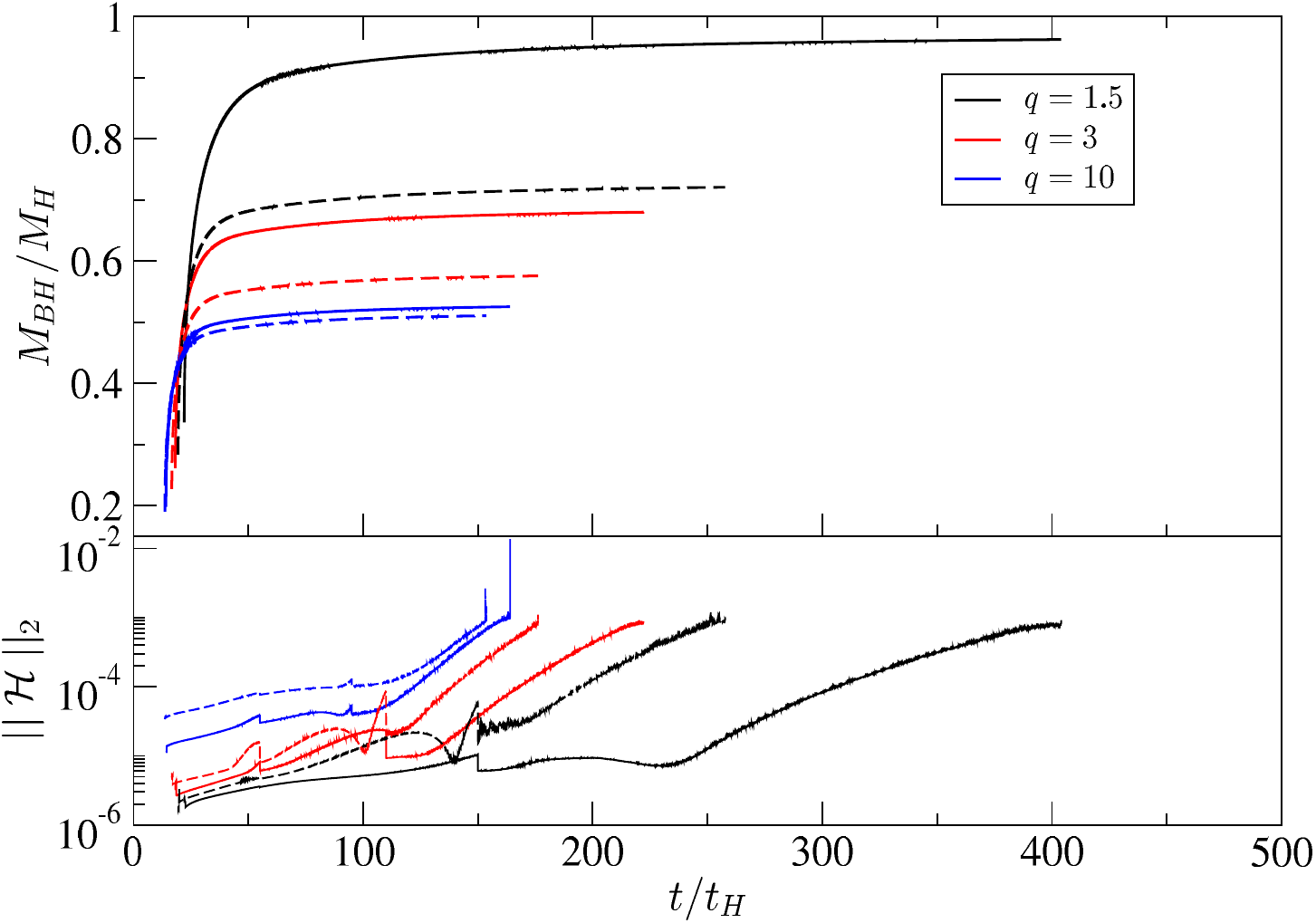} 
\caption{Top: Time evolution of the PBH mass for the profiles $q=1.5$, $q=3$ and $q=10$. Solid line corresponds to Eq.\eqref{basis_pol} and dashed line to Eq.\eqref{eq:lamda} with $\lambda=0$. Bottom: Time evolution of the corresponding  Hamiltonian constraints. In all cases $\bar{\delta}_{m}-\delta_{c}=0.005$.}
\label{fig:acretion_mass}
\end{figure}

Some examples of the time evolution of $M_{BH}(t)$ for different curvature profiles are given in Fig.(\ref{fig:acretion_mass}). 

In \cite{Niemeyer1,musco2007} it was found that for $\bar{\delta}_{m} \approx \delta_{c}$ the following scaling law is satisfied  
\begin{equation}
M_{\rm BH,f} = M_{H} {\cal K} (\bar{\delta}_{m}-\delta_{c})^{\gamma},
\label{eq:scaling}
\end{equation}
where $\gamma \approx 0.36$ during radiation domination where  ${\cal K}$ is a constant whose value depends on the curvature profile.

As  shown explicitly in \cite{escriva_solo}, the scaling law start to be inaccurate for $\bar{\delta}_{m}-\delta_{c}\gtrsim 10^{-2}$, where the profile used was a Gaussian profile  corresponding to  Eq.\eqref{eq:lamda}  with $\lambda=0$ and $q=1$. Here we  consider  different profiles, showing how the constant $\mathcal{K}$ can vary significantly. The value of $\mathcal K$ is important for the estimation of PBH abundance since the latter is proportional to it \cite{germaniprl}. Usually, in the literature it is commonly assumed ${\cal K}\approx O(1)$, but it has not been investigated systematically the dependency of $\mathcal{K}$ on the initial curvature profiles. 

To modulate the existence of a mass excess sufficiently far away from the peak of $\mathcal{C}(r)$, we have used another profile that comes from the junction of two curvatures of Eq.\eqref{basis_pol}, we will refer it as the two-tower profile, and it's expression in terms of the compaction function $\mathcal{C}_{b}$ (refereed to Eq.\eqref{basis_pol}) is directly given by Eq.\eqref{eq:dos_torres},


\begin{equation}
\mathcal{C}_{tt}(r) = \mathcal{C}_{b}(r, \delta_{1},q_{1},r_{m1})+\theta(r-r_{j}) \mathcal{C}_{b}(r-r_{j},\delta_{2},q_{2},r_{m2}).
\label{eq:dos_torres}
\end{equation}
where $\mathcal{C}_{b}$ is equal to
\begin{equation}
\mathcal{C}_{b}(r, \delta_{j},q_{j},r_{mj}) = \delta_{j} \left(\frac{r}{r_{mj}}\right)^{2}\frac{1 + 1/q_{j}}{1+\frac{1}{q_{j}}\left(\frac{r}{r_{mj}}\right)^{2(q_{j}+1)}}; 
\label{eq:dos_torres2}
\end{equation}

\begin{figure}[!htbp]
\centering
\includegraphics[width=1\linewidth]{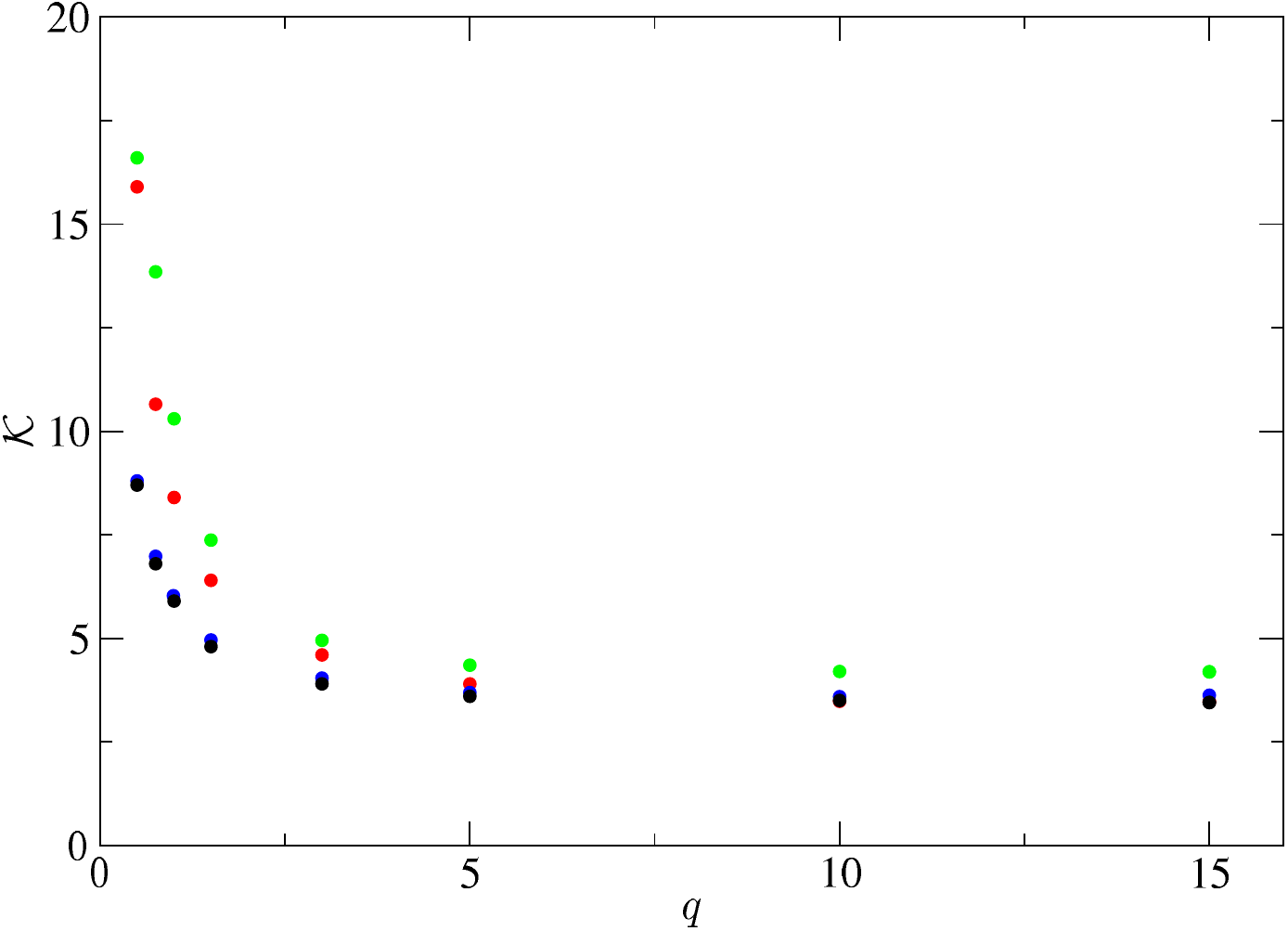} 
\caption{The constant $\cal K$ defined in Eq.\eqref{eq:scaling} is plotted as a function  $q$ for the profiles in Eq.\eqref{basis_pol} (red),  Eq.\eqref{eq:lamda} with $\lambda=0$ (black),   Eq.\eqref{eq:lamda} with $\lambda=1$ (blue) and  Eq.\eqref{eq:dos_torres} (green). The parameters used for the profile $\mathcal{C}_{tt}(r)$ are $q_{2}=3$, $r_{j}=2r_{m1}$, $\mathcal{C}_{tt(\rm peak,2)}=0.3$, and $\delta_{2}$ is obtained from Eq.\eqref{eq:pico2} using the previous values, and $q_{1}=q$.}
\label{fig:K}
\end{figure}

It is shown in Fig.(\ref{fig:profiles}). We consider always that the second peak of $\mathcal{C}$ is lower than the first one at $\delta_{1}$, this ensures the first peak collapse and forms the AH. \footnote{In the situation with $\mathcal{C}_{tt(\rm peak,2)} \geq \delta_{1}$, the second peak could be the dominant contribution for the collapse, therefore the definition of the "threshold" may be different. Although that, we don't consider this situation in this work and we leave this question for future research.} The value of the first peak is directly given by $\delta_{1}$, and the value of the second can be modulated through the following equation:
\begin{equation}
\mathcal{C}_{tt(\rm peak,2)} = \frac{(1+q_{1})(r_{j}+r_{m2})^{2} \delta_{1}}{q_{1} r_{m1}^{2}+(r_{j}+r_{m2})^{2}(r_{j}+r_{m2}/r_{m1})^{2q_{1}}}+\delta_{2}
\label{eq:pico2}
\end{equation}
To obtain the value of $\mathcal{K}$ we have computed $M_{\rm BH,f}$ taking for $10^{-3} < \bar{\delta}_{m}-\delta_{c}(q) < 10^{-2}$ and performed a fit of the formula in Eq.\eqref{eq:scaling}, using  $\gamma = 0.357$ \cite{gamma1,gamma2}.  
In Fig.\ref{fig:K} we show the values of $\cal K$ for different profiles. Contrary to the case of the $\bar{\delta}_{m}$ there can be a substantial difference for the value $\cal K$ computed for different curvature profiles, since the accretion process is affected by the shape of the profile beyond the peak of the compaction function $\mathcal{C}(r)$.

\begin{figure}[!htbp]
\centering
\includegraphics[width=1.0\linewidth]{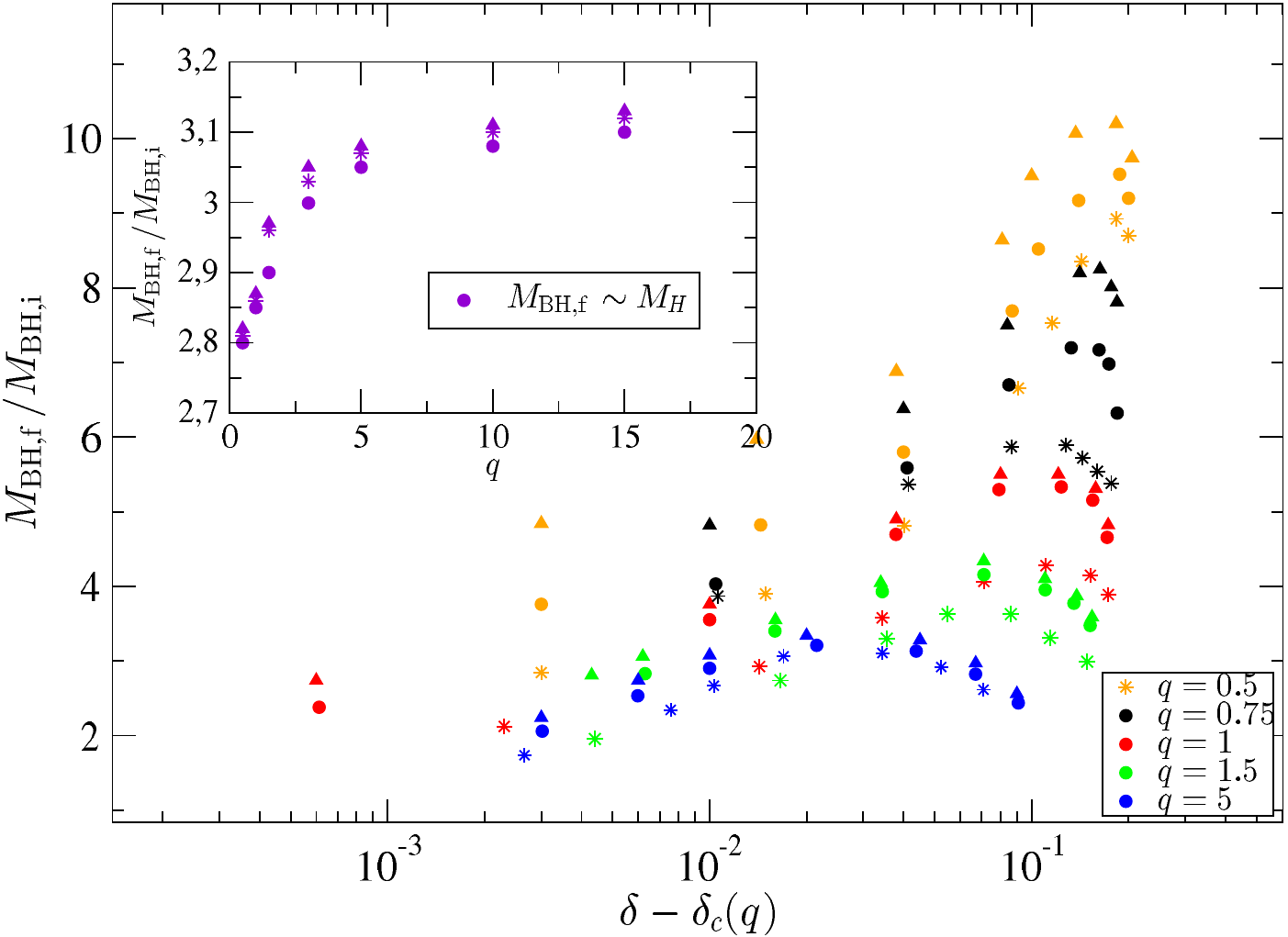} 
\caption{The ratio $M_{\rm BH,f}/M_{\rm BH,i}$ is plotted as a function of $\bar{\delta}_{m}-\delta_{c}(q)$ for different profiles. Circles correspond to Eq.\eqref{basis_pol}, stars to Eq.\eqref{eq:lamda} with $\lambda=0$ and triangles to Eq.\eqref{eq:dos_torres}. The subplot shows the ratio $M_{\rm BH,f}/M_{\rm BH,i}$ for PBHs with $M_{\rm BH,f} \simeq M_{H}$.}
\label{fig:acretion}
\end{figure}

In the case of the profiles given in Eq.\eqref{basis_pol} $\mathcal{K}$ tends  to $\approx 3.5$ for large values of $q$. The value of $\mathcal{K}$ tend to increase as $q$ decreases, as shown Fig.(\ref{fig:K}). 
Numerically we were not able to obtain the final mass $M_{\rm BH,f}$ for profiles $q \lesssim 0.5$, due to conic singularities, as already found in \cite{universal2}. 
As can be see in Fig.(\ref{fig:acretion}) the accretion is more important for large $\bar{\delta}_{m}$. 

Sharp profiles, corresponding to large $q$, have larger pressure gradients and therefore the ratio $M_{\rm BH,f}/M_{\rm BH,i}$ is smaller, even for large $\bar{\delta}_{m}$, since the gradients prevent the accretion.
For low $M_{\rm BH,f}$ the ratio $M_{\rm BH,f}/M_{\rm BH,i}$ should be small, as expected \cite{hawking1,size1}. When $M_{\rm BH,f}\simeq M_{H}$, i.e. for PBHs with higher probability to form, we obtain $M_{\rm BH,f} \simeq 3 M_{H}$. On the other hand as shown in the Fig.(\ref{fig:acretion}), we obtained increasing values of $M_{\rm BH,f}/M_{\rm BH,i}$, for decreasing values of $q$ since pressure gradients are smaller.

\section{Conclusions}\label{sec:conclusions}

We have simulated numerically the formation of spherically symmetric primordial black holes (PBHs) seeded by different families \cite{universal1,universal2} of primordial curvature perturbations profiles in a radiation dominated Universe background, performing for the first time a full numerical study of $t_{AH}$, $M_{\rm BH,f}$ and $M_{\rm BH,i}$. 

The masses $M_{\rm BH,i}$ and $M_{\rm BH,f}$ depend on the full shape of the curvature profile, contrary to $\delta_{c}$, which depends only on the shape around the peak of the compaction function \cite{universal1}. The analytical estimation of the upper bound of the PBH size \cite{size1} has been compared with the numerical results, showing good agreement, except  for profiles with $q \approx 2$, in the limit  $\bar{\delta}_{m}$ approaching $\delta_{\rm max}$. 

We have also obtained for the first time a numerical estimation of the accretion effects for different profiles. For  PBHs with masses relevant for dark matter abundance \cite{germaniprl}, corresponding to $M_{\rm BH,f} \approx M_{H}$,  we obtained $M_{\rm BH,f} \approx 3 M_{\rm BH,i}$.

In the future it would be interesting to study the PBHs formation for different equations of state of the perfect fluid and to considerr non spherically symmetric simulations \cite{new2}. It will also be interesting to  study accretion effects in different systems such as the collapse of Q-balls or oscillons \cite{Cotner:2016cvr,Cotner:2018vug}, and compare with  domains walls or vacuum bubbles \cite{deng,vacum_bubles}.

\begin{acknowledgments}

We thank Jaume Garriga and Cristiano Germani for useful comments and illuminating discussions. AE is supported by the Spanish MECD fellowship FPU15/03583 and by the national FPA2016-76005-C2-2-P grants of the Ministerio de Ciencia y Eduacion.

\end{acknowledgments}

\bibliography{refs.bib}

\end{document}